\documentclass[aps,twocolumn,10pt,aps,amsmath,amssymb,nofootinbib,superscriptaddress,showkeys,showpacs]{revtex4-1}
\usepackage{epsfig} 
\usepackage{amsmath} 
\usepackage{graphicx}
\usepackage{color}
\usepackage{float}
\usepackage{soul,xcolor}
\usepackage[font=scriptsize]{caption}
\usepackage{braket}
\usepackage{bbold}
\usepackage{subfig}
\usepackage{braket}
\usepackage{titlesec}
\usepackage{placeins}

\begin{document}
\title{Quantum correlations and the neutrino mass degeneracy problem}

\title{Quantum correlations and the neutrino mass degeneracy problem.}
\author{Khushboo Dixit}
\email{dixit.1@iitj.ac.in}
\affiliation{Indian Institute of Technology Jodhpur, Jodhpur 342011, India}

\author{Javid Naikoo}
\email{naikoo.1@iitj.ac.in}
\affiliation{Indian Institute of Technology Jodhpur, Jodhpur 342011, India}

\author{Subhashish Banerjee}
\email{subhashish@iitj.ac.in}
\affiliation{Indian Institute of Technology Jodhpur, Jodhpur 342011, India}

\author{Ashutosh Kumar Alok}
\email{akalok@iitj.ac.in}
\affiliation{Indian Institute of Technology Jodhpur, Jodhpur 342011, India}


\begin{abstract}
	Many facets of nonclassicality are probed in the context of three flavour neutrino oscillations including matter effects and CP violation. The analysis is carried out for parameters relevant to two ongoing experiments NO$\nu$A and T2K, and also for the upcoming experiment DUNE. The various quantum correlations turn out to be sensitive to the mass-hierarchy problem in neutrinos. This sensitivity is found to be more prominent in DUNE experiment as compared to NO$\nu$A and T2K experiments. This can be attributed to the large baseline and high energy of the DUNE experiment. Further, we find that to probe these correlations, the neutrino (antineutrino) beam should be preferred if the sign of mass square difference $\Delta_{31}$ turns out to be positive (negative).	
\end{abstract}
\maketitle
\section{Introduction}

Quantum Mechanics has proved to be an incredibly successful theory. Not only have its predictions  been verified with great accuracy, but it has also laid the foundation of new realms of technology, ready to revolutionize the information and communication sectors. Surprisingly, despite all of its success, the question of when does a system behave quantum mechanically rather than classically, still waits for a clear and unambiguous answer. This question becomes important while dealing with the nature of correlations between different subsystems of a composite system. These correlations can be spatial as well as temporal. Some of the widely studied spatial quantum correlations are entanglement \cite{Horodecki2009}, steering \cite{Cavalcanti_Steer}, non-locality \cite{Brunner_Bellnonlo} and quantum discord \cite{OlivierDiscord}. The temporal correlations include Leggett-Garg (LG) \cite{LGoriginal} and LG type inequalities \cite{EmaryArticle}.\par 
The  quantum correlations have been studied  mainly in the optical and electronic systems \cite{aspect1981experimental, tittel1998experimental, lanyon2013experimental, indranilsb}.
Recently such studies have been extended to high energy physics owing to the advancement in various experimental facilities, see for example \cite{Caban:2006ij, bramon2007review, Hiesmayr:2007he, blasone2009entanglement, Blasone:2015lya, nikitin2015proposal, alok2016quantum, banerjee2015quantum, banerjee2016quantum, Naikoo:2017fos, Cervera-Lierta:2017tdt, Kerbikov:2017spv,  Naikoo:2018vug, Richter-Laskowska:2018ikv}. The concept of single particle entanglement has been introduced in previous studies \cite{Zanardi,van,Cunha} which have also been demonstrated experimentally with single photon systems \cite{Lombardi,Lvovsky,Hessmo}. Later, the experimental schemes to probe nonlocality were generalized to include massive particles \cite{Dunningham}. In \cite{blasone2009entanglement}, an experimental scheme is discussed for transferring this form of entanglement to spatially separated modes of stable leptonic particles. It allows to put mode entanglement in neutrino oscillations on equal footing with that in atomic and optical systems. Therefore, different flavour modes of neutrinos can be expressed as legitimate individual entities and entanglement in these flavour modes, i.e. mode-entanglement, studied. An extensive study of quantum correlations in the context of two and three flavour neutrino oscillations is given in \cite{alok2016quantum} and \cite{banerjee2015quantum}, respectively. However, in these works, matter and $CP$ (charge conjugation-parity) violating effects were not taken into account.\par

In this work, we study various facets of nonclassicality, quantified by spatial quantum correlations such as flavour entropy, geometric entanglement, Mermin and Svetlichny inequalities, in the context of three flavour neutrino oscillations, by taking into account the matter effects and $CP$ violation. We discuss the behavior of these quantum correlations for the ongoing experiments like NO$\nu$A  and T2K, and also for the upcoming experiment DUNE. We find that the various witnesses show sensitivity to the mass-hierarchy problem and $CP$ violation in neutrino physics. \par
For a general and physically reliable study of the neutrino oscillation phenomena, one should look in terms of the wave packet approach, i.e. localization effects of production and detection processes should be considered. However, the plane wave approximation also holds good since the oscillation probability obtained with the wave-packet treatment is found to be in consonance with the plane-wave oscillation probability averaged over the Gaussian $L/E$ distribution  \cite{Kim}. Here $L$ and $E$ represent the distance travelled by the neutrino and its energy.

The paper is organized as follows: In Sec. (\ref{dynamics}), we give a brief description of the dynamics of the neutrino oscillation in three flavour case in vacuum and constant matter density. Section (\ref{measures}) is devoted to a brief description of various quantum correlations studied in this work. Section (\ref{results}) gives the results and their discussion. We finally summarize our work in Sec. (\ref{conclusion}).

\section{Neutrino dynamics in vacuum and constant matter density} \label{dynamics}

In this section, we briefly describe the neutrino oscillations in vacuum and in constant matter density. To this aim, consider an arbitrary neutrino state $\ket{\Psi(t)}$ at time $t$, which can be represented either in the flavour basis $\{ \ket{\nu_e}, \ket{\nu_\mu},\ket{\nu_\tau}\}$ or in the mass-basis $\{ \ket{\nu_1}, \ket{\nu_2},\ket{\nu_3}\}$ as:
\begin{equation}
\ket{\Psi (t)} = \sum_{\alpha = e, \mu, \tau} \nu_{\alpha}(t) \ket{\nu_{\alpha}}
=  \sum_{i = 1,2,3} \nu_{i}(t) \ket{\nu_{i}}.
\end{equation}
The coefficients in the two representations are connected by a \textit{unitary} matrix~\cite{banerjee2015quantum, alok2016quantum}
\begin{equation}\label{nualpha}
\nu_\alpha(t) = \sum_{i=1,2,3} U_{\alpha i} \nu_{i}(t).
\end{equation} 
A convenient parametrization for $U$  in terms of mixing angles $\theta_{ij}$ and $CP$ violating phase $\delta$ is given in Eq. (\ref{PMNS}).

\begin{widetext}
	\begin{equation}\label{PMNS}
	U(\theta_{12},\theta_{13},\theta_{23},\delta) = 
	\begin{pmatrix}
	c_{12} c_{13} & s_{12} c_{13} & s_{23} e^{-i \delta} \\ - s_{12}c_{23} - c_{12} s_{23}s_{13} e^{i\delta} & c_{12}c_{23}-s_{12}s_{23}s_{13} e^{i\delta} & s_{23}c_{13} \\ s_{13}s_{23} - c_{12}c_{23}s_{13} e^{i\delta} & -c_{12}s_{23}-s_{12}c_{23}s_{13} e^{i\delta} & c_{23}c_{13}\end{pmatrix},
	\end{equation}
\end{widetext}
	where $c_{ij} = \cos\theta_{ij}$, $s_{ij} = \sin\theta_{ij}$ and $\delta$ is the $CP$ violating phase.

The time evolution of massive states is given by  $\nu_i(t) = e^{-iE_i t} ~\nu_i (0)$, which, along-with Eq. (\ref{nualpha}), gives
\begin{equation}
{\nu}_\alpha(t) = U_{f} ~\nu_\alpha (0). \label{flavourevolution}
\end{equation}
Here, $U_f$ is the flavour evolution matrix, taking a flavour state from time $t=0$ to some later time $t$. In matrix form
\begin{equation}\label{Uflv}
\begin{pmatrix}
\nu_e(t) \\ \nu_\mu(t) \\ \nu_\tau(t)
\end{pmatrix}= \begin{pmatrix}
a(t) & d(t) & g(t) \\ b(t) & e(t) & h(t) \\ c(t) & f(t) & k(t)\end{pmatrix} 				\begin{pmatrix}
\nu_e(0) \\ \nu_\mu(0) \\ \nu_\tau(0)
\end{pmatrix}.
\end{equation}

If the state at time $t = 0$ is $\ket{\nu_e}$, then $\nu_\alpha(0) = \delta_{\alpha e}$ ($\alpha =e,\mu,\tau$). Therefore after time $t$, we have $\nu_e (t) = a(t)$, $\nu_{\mu}(t) = b(t)$ and $\nu_{\tau}(t) =  c(t)$. Hence, the wave function can be written as
\begin{equation}\label{nu-e}
\ket{\Psi_e (t)} = a(t) \ket{\nu_e} + b(t) \ket{\nu_{\mu}} + c(t) \ket{\nu_{\tau}}.
\end{equation}
The survival probability is then given by $| \langle \nu_e | \Psi_e (t)  \rangle|^2 =|a(t)|^2$. Similarly, $|b(t)|^2$ and $|c(t)|^2$ are the transition probabilities to $\mu$ and $\tau$ flavour, respectively. The survival and transition probabilities are functions of energy difference $\Delta E_{ij} = E_i - E_j$  ($j,k = 1,2,3$). Also, in the   ultra-relativistic limit, following  standard approximations are adopted:
	\begin{equation}\label{URA}
\Delta E_{ij} \approxeq \frac{\Delta m^2_{ij}}{2 E}; \quad E \equiv |\vec{P}|; \quad t \equiv L.
	\end{equation}
These approximations are quite reasonable in the context of the experiments considered here, since the neutrinos are ultra relativistic with neutrino-masses of the order of a few electron-volts (eV) and the  energy higher than $10^6 ~{\rm eV}$, as discussed in Sec. (\ref{dynamics}) under \textit{neutrino experiments.}\par
\textit{--Occupation number representation:} Given the above formalism, one can introduce the occupation number associated with a given flavour or mass mode \cite{blasone2009entanglement,blasone2014field, banerjee2015quantum}
\begin{equation}
\renewcommand{\arraystretch}{1.0}
\left.\begin{array}{r@{\;}l}
\ket{\nu_e} \equiv \ket{1}_e \ket{0}_\mu \ket{0}_\tau \\\\
\ket{\nu_\mu} \equiv \ket{0}_e \ket{1}_\mu \ket{0}_\tau\\\\
\ket{\nu_\tau} \equiv \ket{0}_e \ket{0}_\mu \ket{1}_\tau
\end{array}\right\} \rm flavour ~~modes \label{occupflv}
\end{equation}
\begin{equation}
\renewcommand{\arraystretch}{1.0}
\left.\begin{array}{r@{\;}l}
\ket{\nu_1} \equiv \ket{1}_1 \ket{0}_2 \ket{0}_3 \\\\
\ket{\nu_2} \equiv \ket{0}_1 \ket{1}_2 \ket{0}_3\\\\
\ket{\nu_3} \equiv \ket{0}_1 \ket{0}_2 \ket{1}_3
\end{array}\right\} \rm  massive ~~modes \label{occupmass}
\end{equation}
Here, $\ket{n}_\alpha$ represents the $n$-th occupation number state of a neutrino in mode $\alpha$.
\begin{equation}\label{nu-e-occup}
\ket{\Psi_e (t)} = a(t) \ket{100} + b(t) \ket{010} + c(t) \ket{001}.
\end{equation}
 Thus the time evolved flavour state (Eq. (\ref{nu-e})) can be viewed as an entangled superposition of flavour modes (Eq. (\ref{nu-e-occup})) with the time dependent coefficients given by Eq. (\ref{Uflv}).
  Care should be taken in dealing with the above defined Fock representations in the flavour and mass basis as they are unitarily inequivalent in the quantum field theoretic description of neutrino oscillations \cite{BLASONE1995283}. Specifically, the unitary equivalence of the flavour and the mass state given in Eq. (\ref{nualpha}), is not valid under the infinite volume approximation as  the flavour and mass eigenstates become orthogonal and the vacuum for definite flavour neutrinos can not be identified with the vacuum state for definite mass neutrinos. However, in this work, we stick to ultra relativist approximation, Eq. (\ref{URA}), under which  the unitary equivalence holds and talk about the various nonclassical witnesses viz., entanglement existing among different flavour modes in a single particle setting. It would be interesting to investigate the behavior of these witnesses by incorporating the various non trivial effects arising from quantum field theoretic treatment of neutrino oscillation viz, vacuum condensation.\par
\textit{--Matter effect:} The matter density has significant effect on the neutrino energy spectrum. The effects of earth's matter density on neutrino oscillations has been studied using various models for matter densities \cite{VKErmilova, krastev1989, petcov1998, akhmedov1999kh, freund2000matter, freund2000m, KD}. To incorporate the matter effect, we are going to use the formalism developed in \cite{ohlsson2000three, OHLSSON2000153}. In vacuum, the Hamiltonian $H_m$ is given by $H_m =  diag[E_1, E_2, E_3]$, where $E_a = \sqrt{m^{2}_{a} + p^2}$, $a=1, 2, 3$ are the energies of the neutrino mass eigenstates $\ket{\nu_a}$, with masses $m_a$ and momentum $p$. When neutrinos propagate through ordinary matter, the Hamiltonian picks up an additional term as a consequence of the weak interaction with the electrons in the matter. This additional potential term is diagonal in the flavour basis and is given by $V_f=  diag[A, 0, 0]$, where $A = \pm \sqrt{2} G_f N_e$ is the matter density parameter and $G_f$ and $N_e$ are the Fermi coupling constant and electron number density, respectively. The sign of the matter density parameter is positive for neutrinos and negative for antineutrinos. We assume that the electron density $N_e$ is constant throughout the matter in which the neutrinos are propagating. In the mass basis, the additional potential term becomes $V_m = U^{-1} V_f U$, where $U$ is given in Eq. (\ref{PMNS}). Thus the Hamiltonian in mass basis is given by  $\mathcal{H}_m = H_m + U^{-1} V_f U$. After some algebra, one finally obtains the matter counterpart of the  flavour evolution matrix defined in Eq. (\ref{flavourevolution}):
\begin{equation}\label{Umatter}
U_{f}(L) = \phi ‎‎\sum_{n=1}^{3} e^{-i \lambda_{n} L} \frac{1}{3\lambda_{n}^{2} + c_1} \left[ (\lambda_{n}^{2} + c_{1}) \mathbf{I} + \lambda_n \tilde{\mathbf{T}} + \tilde{\mathbf{T}}^2 \right].
\end{equation}
Here $\phi \equiv e^{−i L tr H_m/3}$, $\lambda_n ~(n=1, 2, 3)$ are the eigenvalues of $\mathbf{T}$ matrix defined further in Eq. (\ref{Tmatrix}), $\tilde{\mathbf{T}} = U \mathbf{T} U^{-1}$ and $c_1$ = $det\mathbf{T} \times Tr\mathbf{T}^{-1}$. For a multilayer model potential with density parameters $A_1, A_2, A_3\dots A_m$, and lengths $L_1, L_2, L_3\dots L_m$, the net flavour evolution operator will be the product of the operators corresponding to the each density, that is, $U_f|_{Net} = U_f(L_1).U_f(L_2).U_f(L_3)\dots U_f(L_m)$.

\begin{widetext}
	\begin{equation}\label{Tmatrix}
	\mathbf{T} =\begin{pmatrix}
	A U_{e1}^{2} - \frac{1}{3} A + \frac{1}{3} ( E_{12} + E_{13} ) & A U_{e1} U_{e2} & A U_{e1} U_{e3} \\ A U_{e1} U_{e2} & A  U_{e2}^{2} - \frac{1}{3} A + \frac{1}{3} ( E_{21} + E_{23} ) & A U_{e2} U_{e3}   \\ A U_{e1} U_{e3}  &  A U_{e2} U_{e3}  & A U_{e3}^{2} - \frac{1}{3} A + \frac{1}{3}(E_{31} + E_{32})
	\end{pmatrix}.
	\end{equation}
\end{widetext}

\textit{--Neutrino experiments:}
\begin{itemize}
	\item T2K (Tokai-to-Kamioka) is an off-axis experiment \cite{abe2014precise,abe2013measurement} using a $\nu_{\mu}$-- neutrino beam originating at J-PARC (Japan Proton Accelerator Complex) with energy-range of approximately 100 MeV to 1 GeV and the baseline of 295 km.
	
	\item NO$\nu$A (NuMI Off-Axis $\nu_e$ Appearance), the long baseline experiment, uses neutrinos from NuMI (Neutrinos at the Main Injector)  beamline at Fermilab optimized to observe $\nu_\mu \rightarrow \nu_e$ oscillations. This experiment uses two detectors, both located at 14 mrad off the axis of the NuMI beamline, the near and far detectors are located at 1 km and  810 km from the source, respectively. The flavour composition of the beam is $92.9\%$ of  $\nu_\mu$  and $5.8\%$  of $\bar{\nu}_\mu$  and $1.3\%$ of $\nu_e$ and $\bar{\nu}_e$; the energy of the neutrino beam varies from 1.5 GeV to 4 GeV. The spectrum for NuMI beamline for various off-axis locations is given in \cite{patterson2013nova, muether2013nova, adamson2016first}.

	\item DUNE is an experimental facility which uses NuMI neutrino beam with energy range of 1 - 10 GeV from Fermilab and has a long baseline of 1300 km. This enables L/E, of about $10^3$ km/GeV, to  reach good sensitivity for $CP$ measurement and determination of mass hierarchy \cite{DUNE}.
\end{itemize}
The matter density in all these experiments is approximately $2.8 gm/cc$, which corresponds to the density parameter $A \approx 1.01 \times 10^{-13} eV$.\par
In the next section we analyze the behavior of various quantum correlations in the context of the experiments described above.

\section{Measures of Quantum Correlations}\label{measures}

The  general form of Eq. (\ref{nu-e}), for initial state $\ket{\nu_\alpha}$, can be written as:
\begin{equation}\label{psialpha}
\ket{\Psi_{\alpha} (t)} = \xi_1(t) \ket{\nu_e} + \xi_2(t) \ket{\nu_{\mu}} + \xi_3(t) \ket{\nu_{\tau}}. 
\end{equation}
with
\begin{equation}
\begin{cases}
\vspace{2mm}
\xi_1(t) = a(t), ~\xi_2(t) = b(t), ~\xi_3(t) = c(t) ,& \text{if } \alpha = e\\ 		\vspace{2mm}
\xi_1(t) = d(t), ~\xi_2(t) = e(t), ~\xi_3(t) = f(t) ,& \text{if } \alpha = \mu\\
\xi_1(t) = g(t), ~\xi_2(t) = h(t), ~\xi_3(t) = k(t) ,& \text{if } \alpha =\tau
\end{cases}
\end{equation}
where $a(t),~ b(t), ~c(t)~ .~ .~ .~ k(t)$  are the elements of $U_f$ matrix defined in Eq.(\ref{flavourevolution}) for vacuum. In matter, the corresponding elements of the flavour evolution of matrix (\ref{Umatter}), are used. 
Equivalently, Eq. (\ref{psialpha}) can be written in the occupation number representation as:
\begin{equation}\label{psi}
\ket{\Psi_{\alpha} (t)} = \xi_1(t) \ket{100} + \xi_2(t) \ket{010} + \xi_3(t) \ket{001}.
\end{equation}

\begin{figure*}[ht] 
	\centering
	\includegraphics[width=76mm]{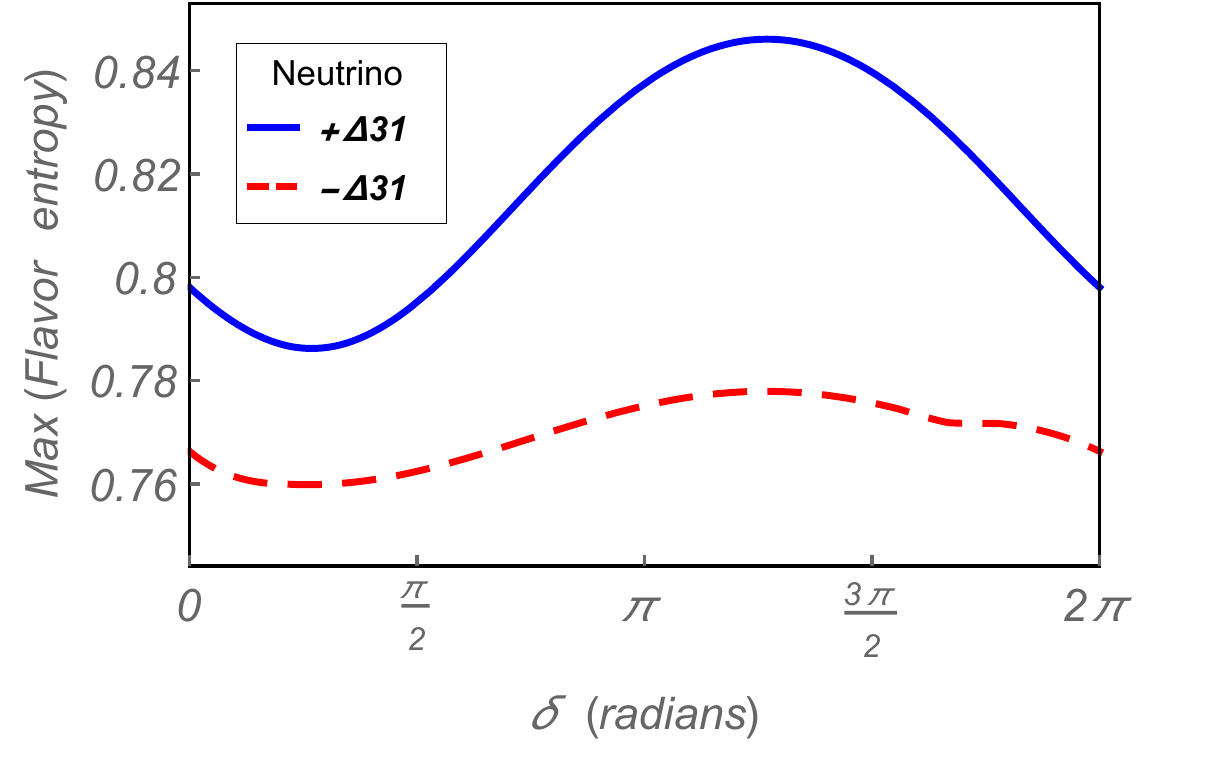}
	\includegraphics[width=76mm]{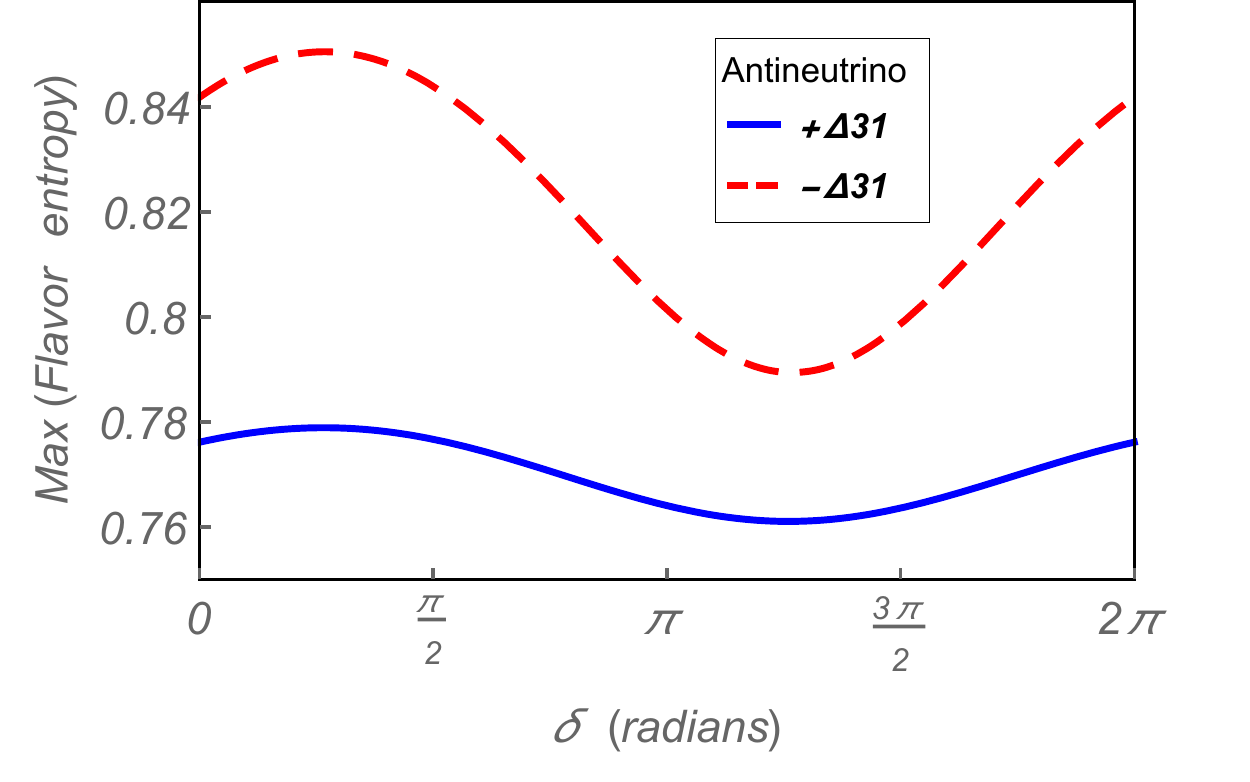}\\
	\includegraphics[width=76mm]{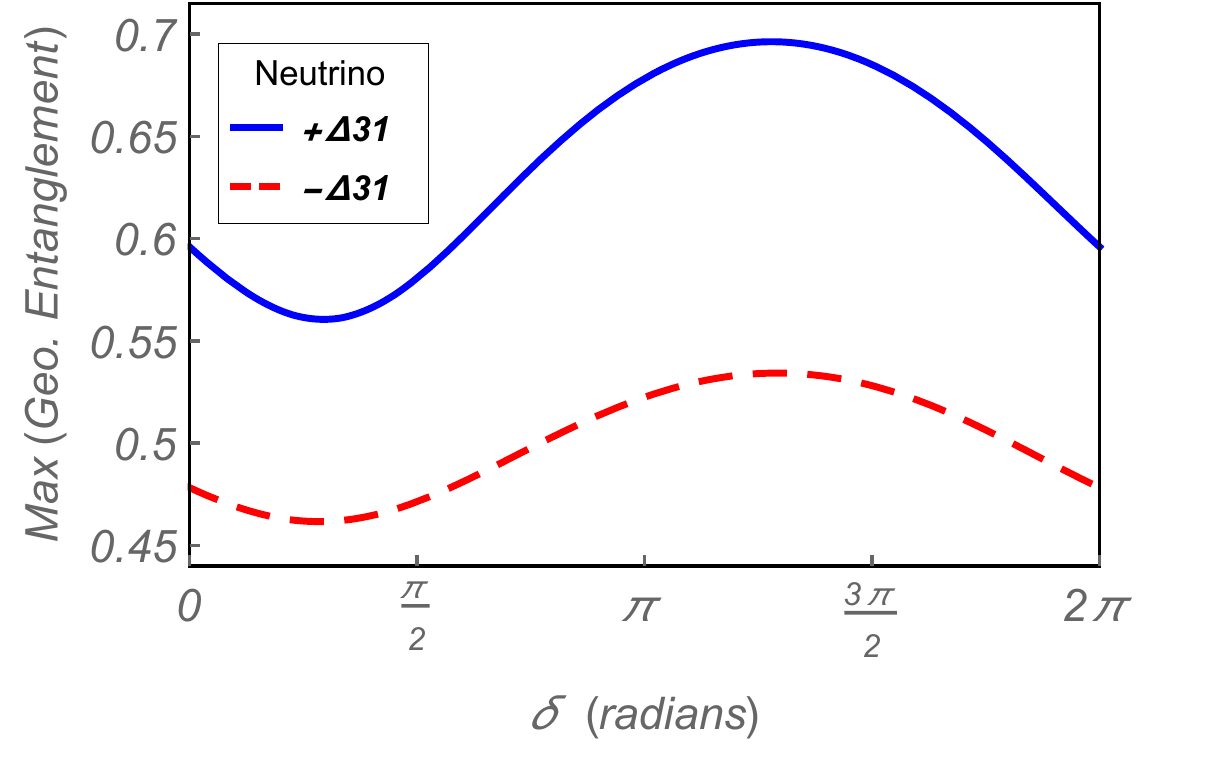}
	\includegraphics[width=76mm]{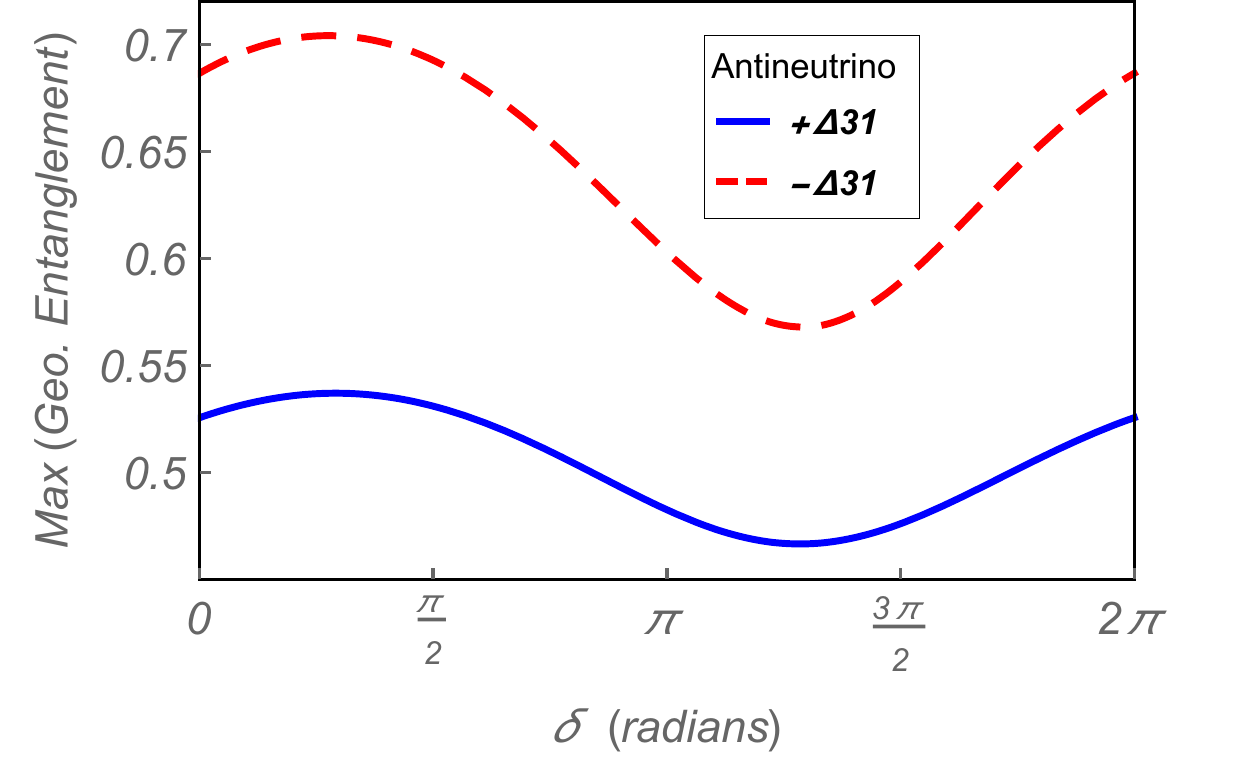}\\
	\includegraphics[width=76mm]{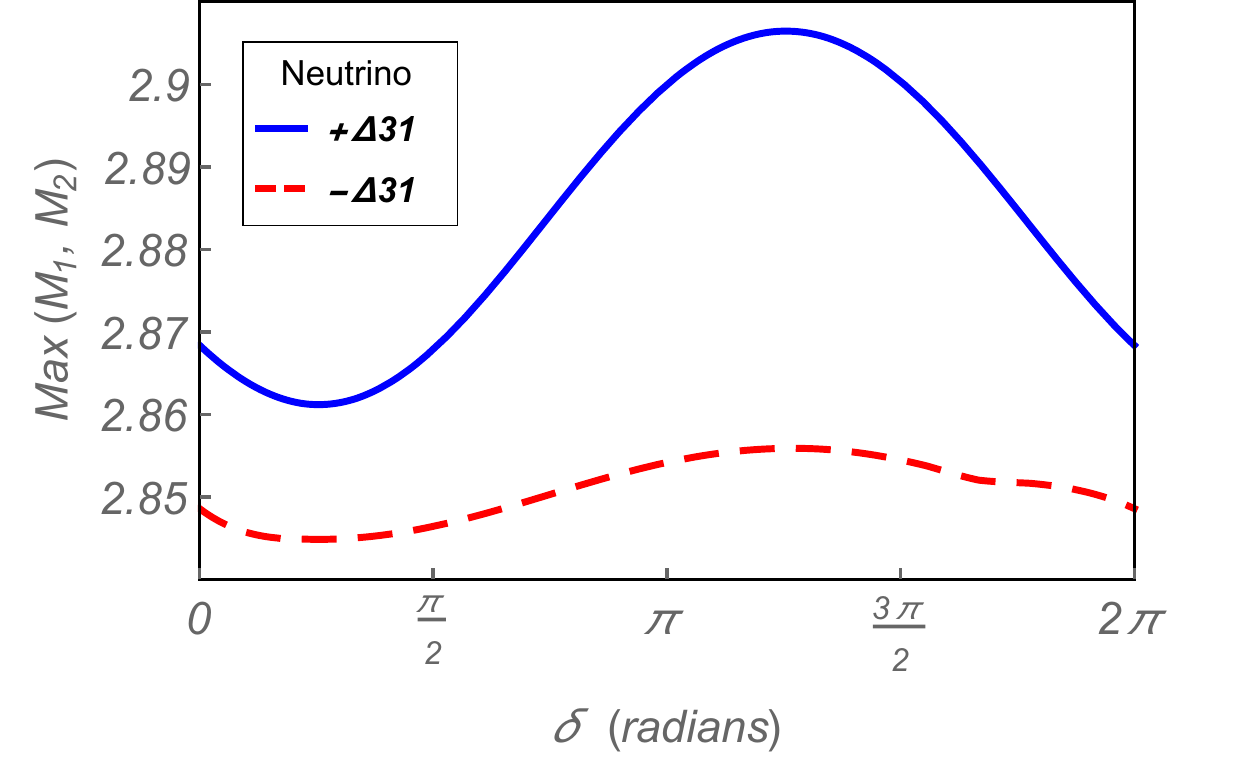}
	\includegraphics[width=76mm]{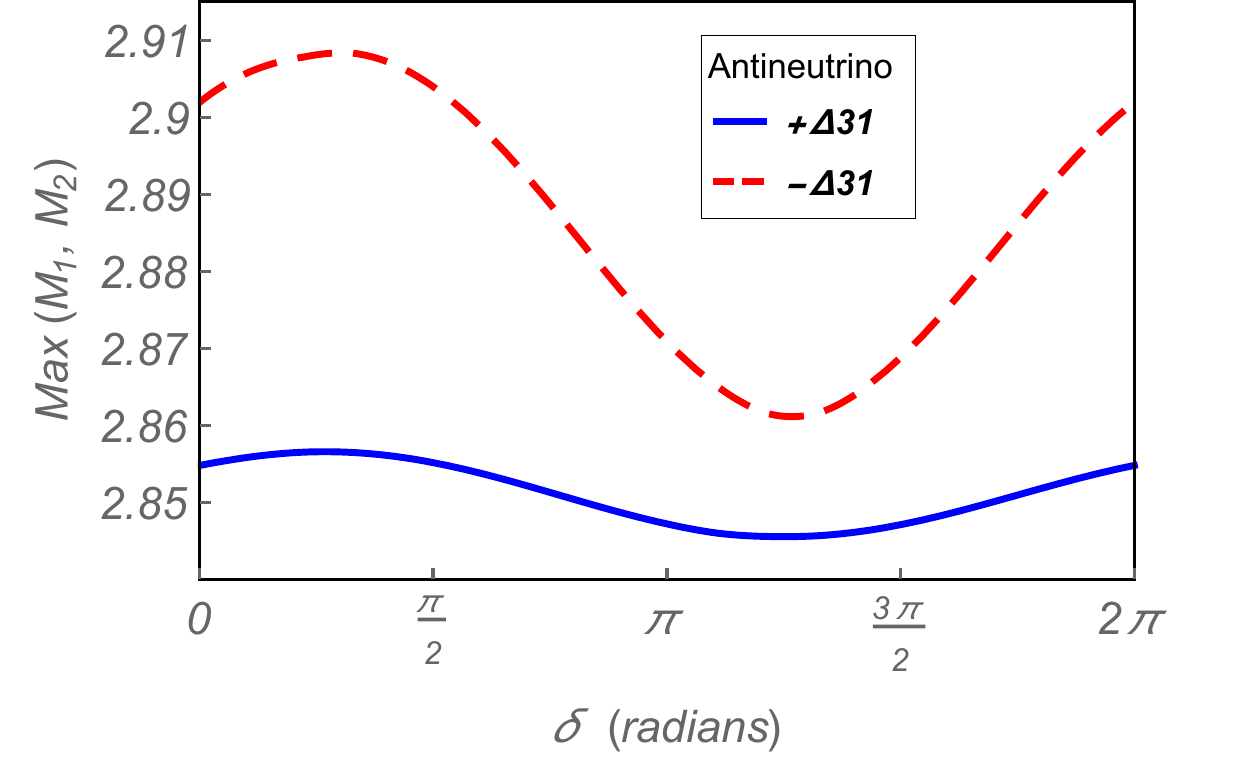}\\
	\includegraphics[width=76mm]{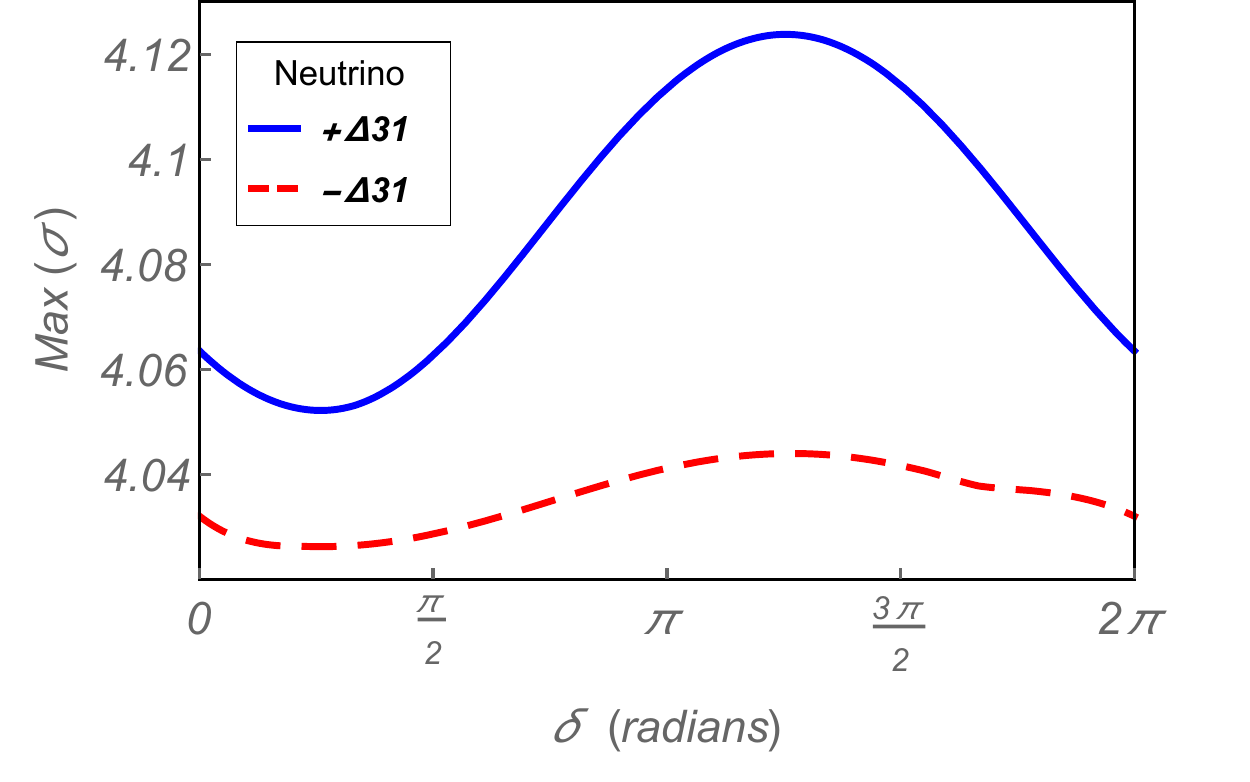}
	\includegraphics[width=76mm]{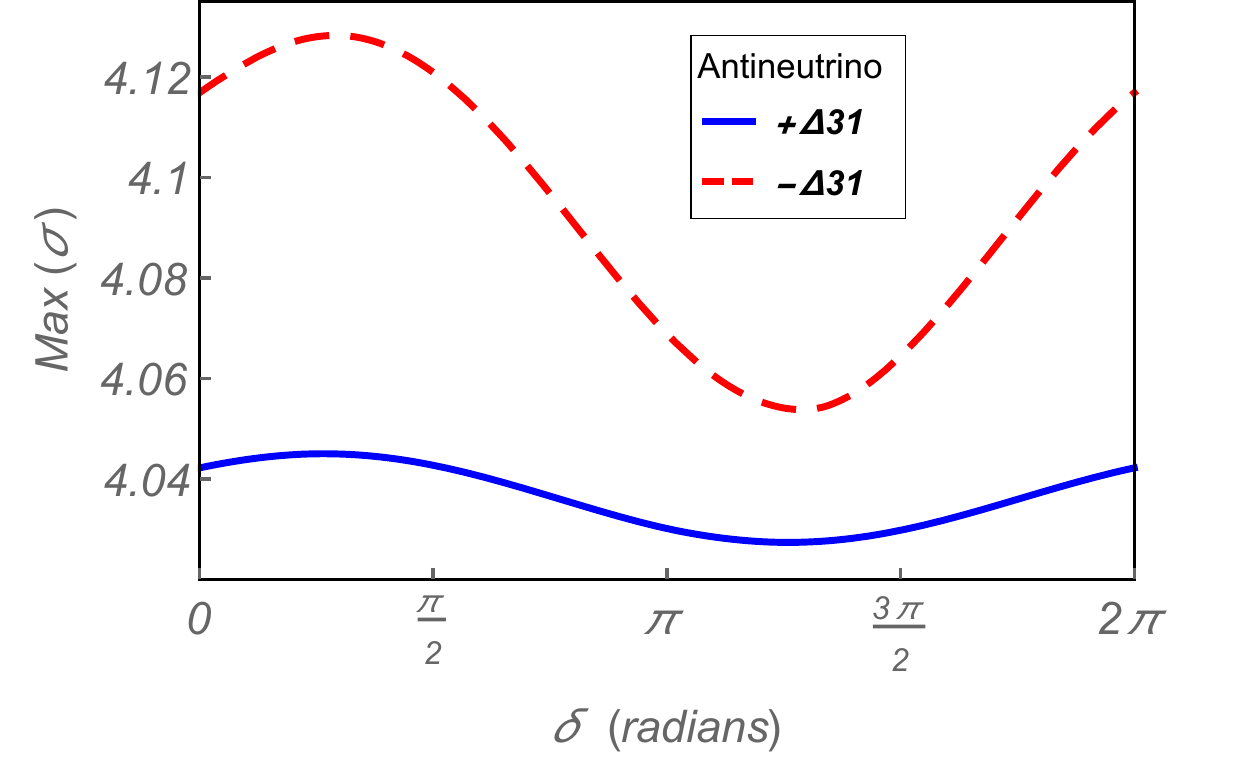}
	\caption{DUNE. The maximum of various quantum correlations  such as Flavor entropy (First row), Geometric entanglement (second row), Mermin parameters ($M_1, M_2$) (third row) and Svetlichny parameter ($\sigma$) (fourth row) depicted with respect to the $CP$ violating phase $\delta$ for DUNE experiment. The left and right panels pertain to the neutrino and antineutrino case, respectively. Solid(blue) and dashed(red) curves correspond to the positive and negative signs of $\Delta_{31}$, respectively. The mixing angles and the squared mass differences used are $\theta_{12} = 33.48^o$, $\theta_{23} = 42.3^o$, $\theta_{13} = 8.5^o$, $\Delta_{21} = 7.5\times 10^{-5} eV^2$, $\Delta_{32} \approx \Delta_{31} = 2.457 \times 10^{-3} eV^2$. The energy range used is $E: 1-10GeV$ and the baseline used is 1300 km. The neutrinos pass through a matter density of $2.8gm/cc$.  }
	\label{DUNE}
\end{figure*}
\FloatBarrier
With this general setting, we now discuss various facets of quantum correlation
	\begin{enumerate}
	\item \textit{Flavor Entropy}:
	For the pure states (\ref{psi}), the standard measure of entanglement is given as \cite{banerjee2015quantum}
	\small
	\begin{align}
	\mathcal{S}(|\xi_{i}|^2) &= -\sum_{i=1}^{3} |\xi_i |^2 \log_2(|\xi_i|^2) \nonumber \\& -\sum_{i=1}^{3} (1-|\xi_i |^2) \log_2(1-|\xi_i|^2). \label{flavourentropy}
	\end{align}
	\normalsize
	This measure serves as a tool to probe the nonclassicality of the system. In the context of neutrino oscillation, the flavour entropy parameter $\mathcal{S} = 0$ for an initially prepared neutrino state $\nu_\alpha$ ($\alpha=e, \mu,\tau$), and reaches its upper bound $\mathcal{S}=1$ for the maximally nonclassical state in the $W$ class $\frac{1}{\sqrt{3}} (\ket{100} + \ket{010} + \ket{001})$\cite{cereceda2002three}.\par

	\item \textit{Tripartite geometric entanglement}:
	Tripartite geometric entanglement $G$ for the pure states, given in Eq. (\ref{psi}), is defined as the cube of the geometric mean of Shannon entropy over every bipartite section.
	\begin{equation}\label{GeoEnt}
	G = H(\xi_1(t)^2)H(\xi_2(t)^2)H(\xi_3(t)^2), 
	\end{equation}  
	where $  H(p)\equiv - p ~log_2(p)-(1-p)~log_2(1-p)$ is the bipartite entropy. This is a weaker condition than genuine tripartite nonlocality discussed below. The genuine tripartite entanglement does not exist if $G=0$.  
	
	\item \textit{Absolute and genuine tripartite nonlocality (Mermin and Svetlichny inequalities)}: 
	The violation of a Bell type inequality (viz., CHSH) for a two qubit state is said to imply nonlocality. A generalization to three party system is not straightforward. Mermin inequality is based on the assumptions that all the three qubits are locally and realistically correlated; hence a violation would be a signature of the tripartite nonlocality shared among the qubits. It was shown in \cite{Collins, Scarani} that the biseparable states also violate the Mermin inequality. This motivated Svetlichny to formulate a hybrid nonlocal-local realism based inequality, the Svetlichny inequality. A three qubit system may be nonlocal if nonclassical correlations exist between two of the three qubits. Such a state would be absolute nonlocal and will violate Mermin inequality \cite{mermin1990extreme} for a particular set of detector setting ($A$,$B$,$C$) and ($A^\prime$,$B^\prime$,$C^\prime$). The two Mermin inequalities are: 
	\small
	\begin{align}
	M_1 &\equiv \left< ABC'+ AB'C + A'BC - A'B'C'\right> \leq 2,\nonumber\\
	M_2 &\equiv \left< ABC - A'B'C - A'BC' - AB'C'\right> \leq 2. \label{mermin}
	\end{align} 
	\normalsize
	However, a violation of Mermin inequality does not necessarily imply genuine tripartite nonlocality. A state violating a Mermin inequality may fail to violate a Svetlichny inequality, which provides a sufficient condition for genuine tripartite nonlocality \cite{svetlichny1987distinguishing} and
	is given by
	\begin{equation}
	\sigma \equiv M_1 + M_2 \leq 4. \label{svetlichny}
	\end{equation}
	\begin{figure}[h!] 
		\centering
		\includegraphics[width=75mm]{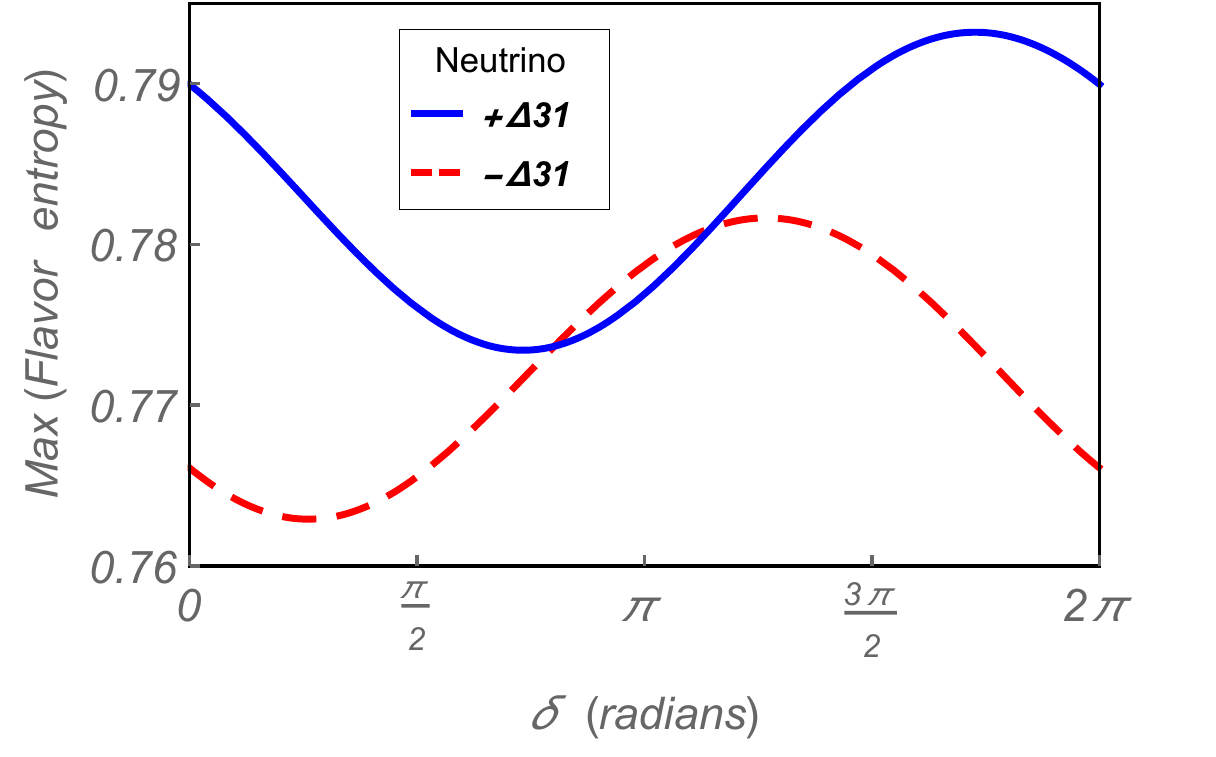}\\
		\includegraphics[width=75mm]{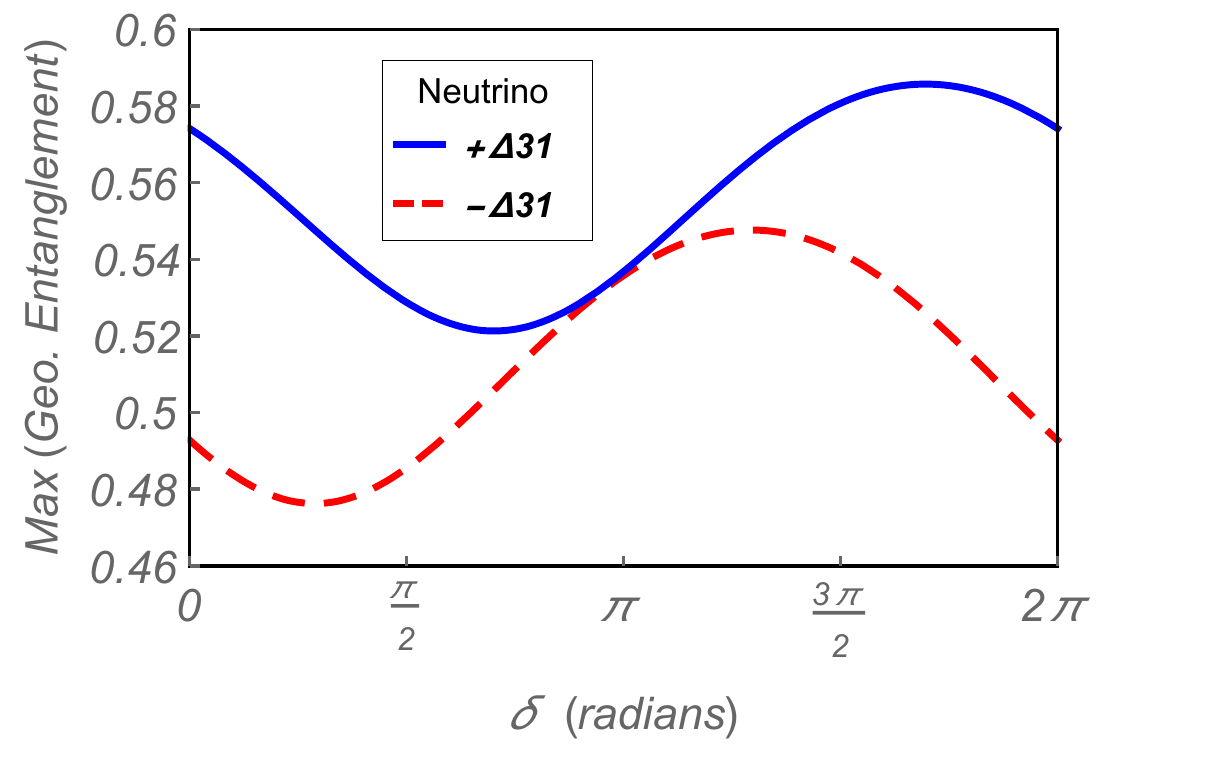}\\
		\includegraphics[width=75mm]{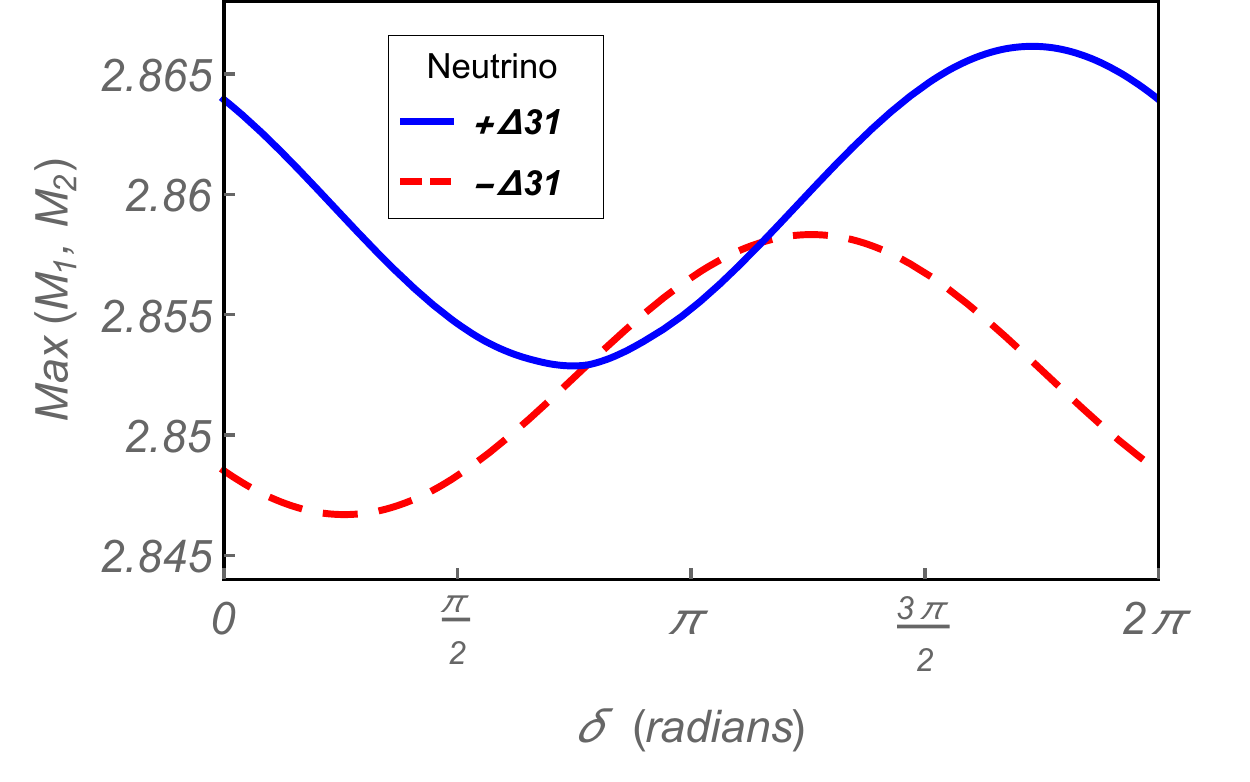}\\
		\includegraphics[width=75mm]{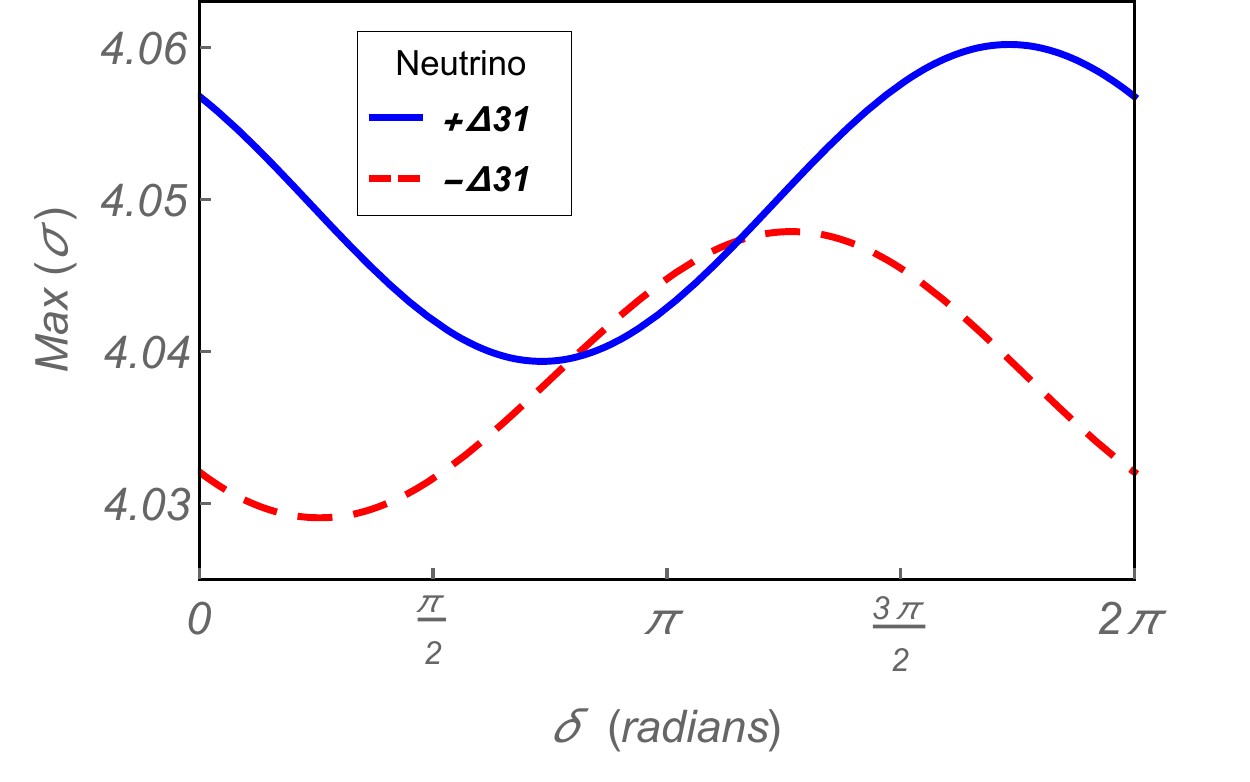}
		\caption{NO$\nu$A: Quantum correlations such as Flavor entropy (first), Geometric entanglement (second), Mermin parameter ($M_1, M_2$) (third) and Sevtlichny parameter ($\sigma$) (fourth) parameters, plotted with respect to the $CP$ violating phase $\delta$ for NO$\nu$A experiment for the case of neutrinos. The energy is varied between $1.5-4GeV$ and the baseline is chosen as 810 km. The various mixing angles and squared mass differences used are the same as for Fig(\ref{DUNE}).}
		\label{NOvA}
	\end{figure}
\end{enumerate}

\begin{figure}[h!] 
	\centering
	\includegraphics[width=75mm]{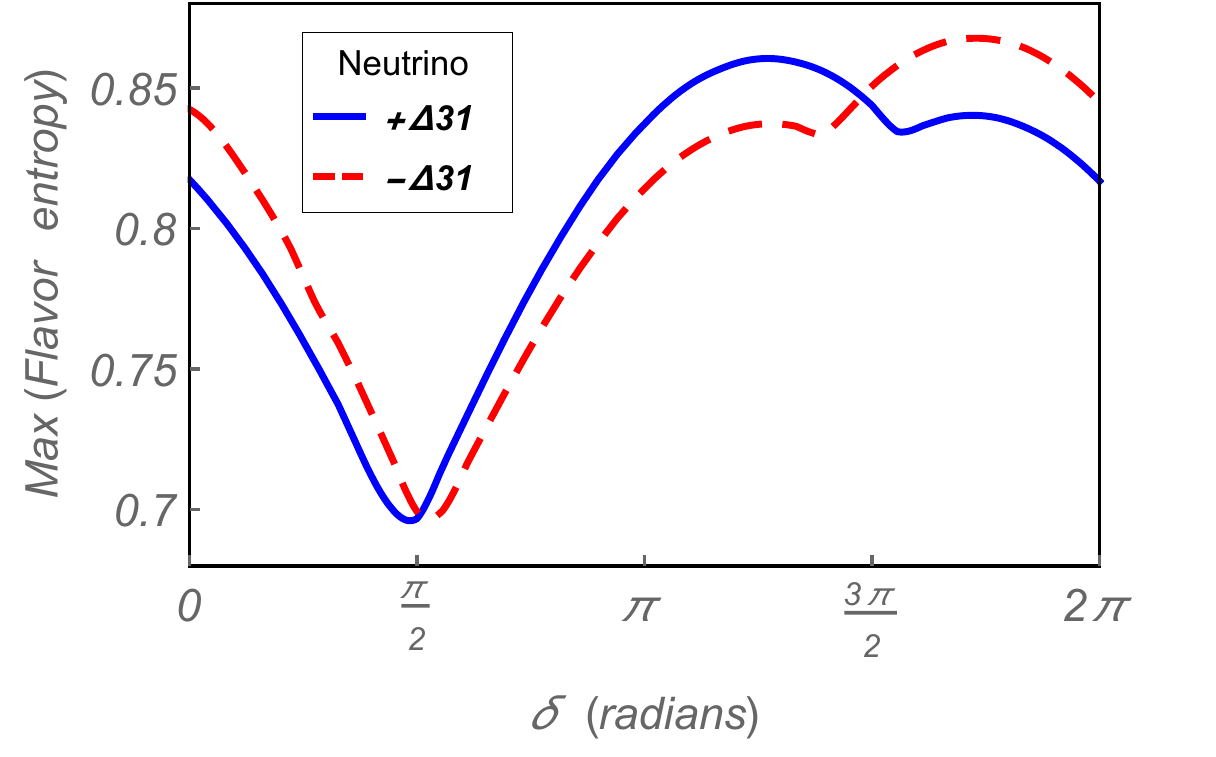}\\
	\includegraphics[width=75mm]{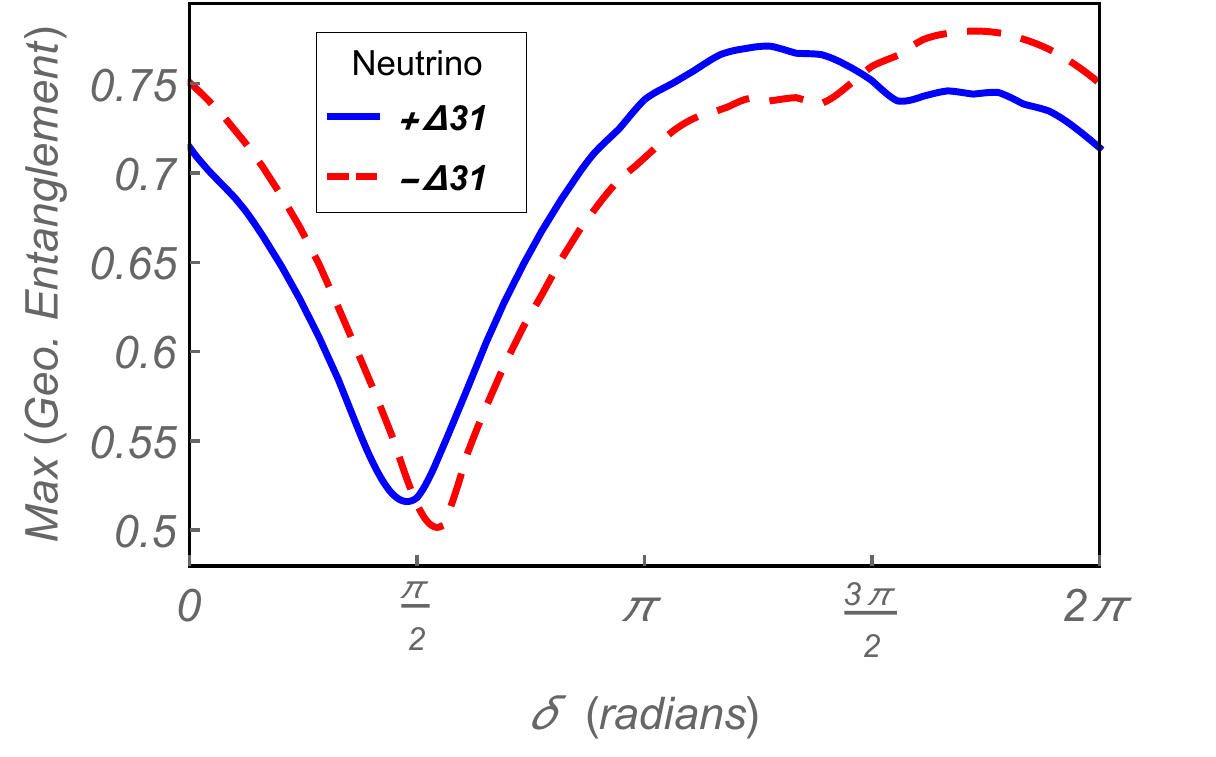}\\
	\includegraphics[width=75mm]{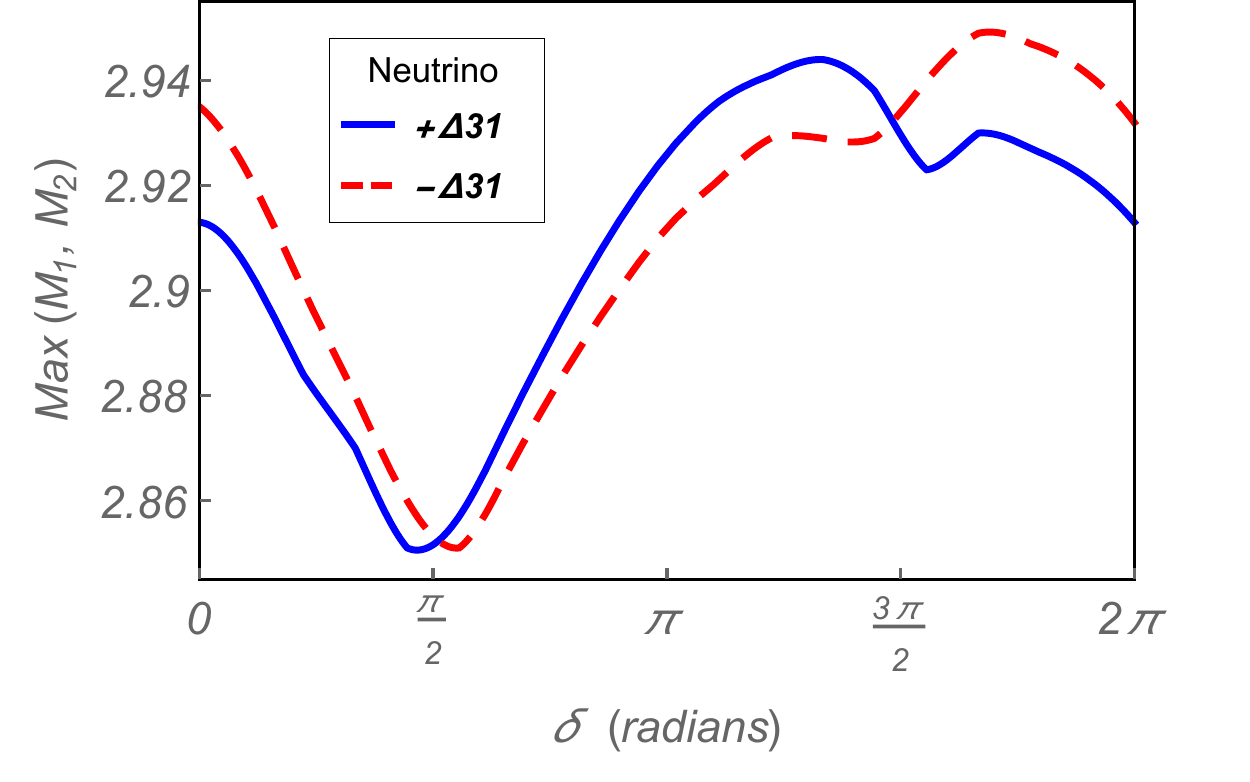}\\
	\includegraphics[width=75mm]{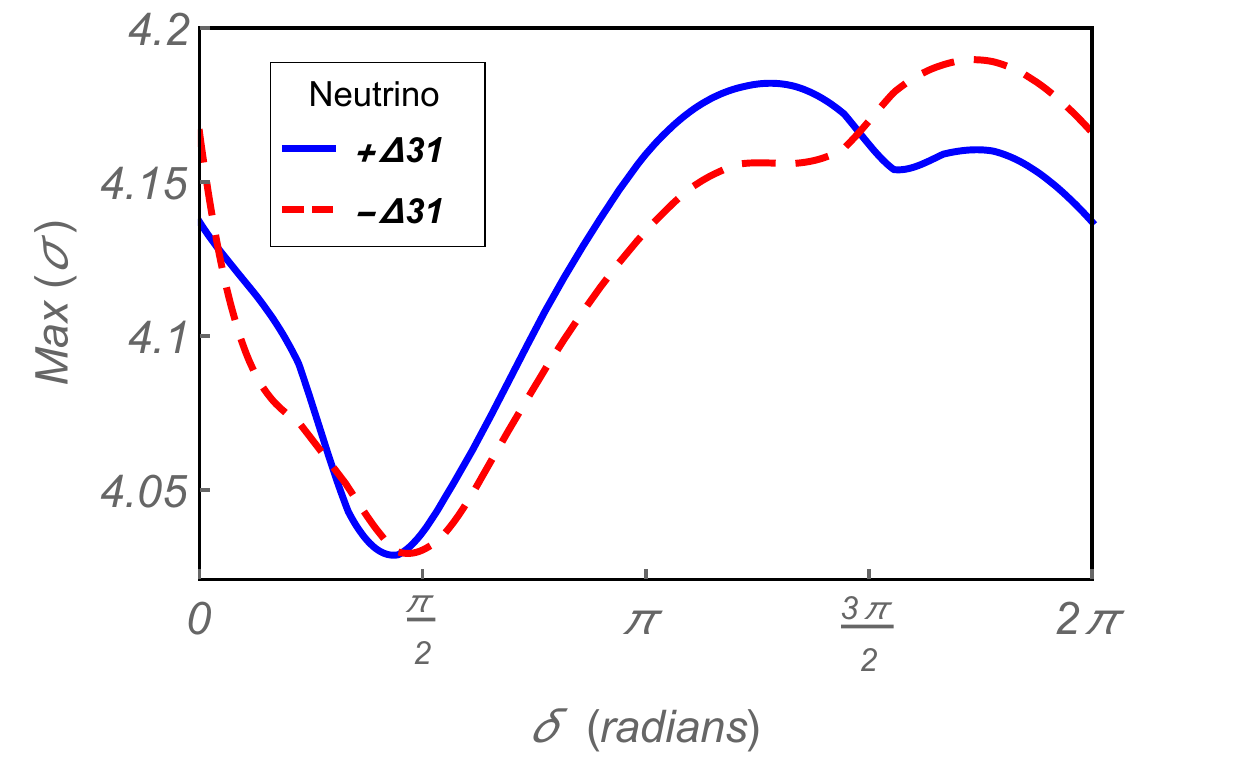}
	\caption{T2K: Showing  Flavor entropy (first), Geometric entanglement (second), Mermin parameter ($M_1, M_2$) (third) and Sevtlichny parameter $(\sigma)$ (fourth) parameters, as function of the $CP$ violating phase $\delta$. The energy is taken between $0.1 - 1 GeV$ and the baseline is 295 km. }
	\label{T2K}
\end{figure}

\section{Results and Discussion}\label{results}

For DUNE experiment, Fig. (\ref{DUNE}) depicts the variation of the maximum of various quantum witnesses like flavour entropy, geometric entanglement, Mermin parameters ($M_1$, $M_2$) and Svetlichny parameter ($\sigma$) with respect to the $CP$ violating phase $\delta$,  for the case of neutrino and antineutrino, respectively. It can be seen that all the witnesses show different characteristics for the positive and negative signs of large mass square difference $\Delta_{31}$. Figures (\ref{NOvA}) and (\ref{T2K}) depict the same for ongoing NO$\nu$A and T2K experiments, for neutrino beam. The  corresponding antineutrino plots show  similar features, such as inversion of mass-hierarchy, as in the DUNE plots and hence are not depicted here.\par
A general feature observed in these results is that the different measures of nonclassicality are sensitive to the sign of $\Delta_{31}$. The distinction being more prominent in DUNE experiment compared to the NO$\nu$A and T2K experiments. This can be attributed to the high energy and long baseline of the DUNE experiment. 

The quantum correlation measures studied in this work can attain their upper bounds for some specific values of $L/E$ \cite{banerjee2015quantum}. In the present study, however, by taking into account the matter effects and CP violation, we are restricting $L/E$ within the experimentally allowed range; consequently the various nonclassical measures do not reach their maximum allowed values.
Mermin inequalities are violated for all values of $\delta$ which means that if one of the three parties is traced out, still there will be residual nonlocality in the system. Violation of the Svetlichny inequality reflects the nonlocal correlation between every subsystem of the tripartite system. To achieve significant violation of correlation measures one should use neutrino-beam if the sign of $\Delta_{31}$ is positive (normal mass hierarchy), while antineutrino-beam should be used in case of negative sign of $\Delta_{31}$ (inverted mass hierarchy).\par
 From the definitions of flavour entropy (Eq. (\ref{flavourentropy})) and geometric entanglement (Eq. (\ref{GeoEnt})), it is clear that these are measurable quantities since these are written in terms of survival and oscillation probabilities making them suitable for  experimental verification. Expressing the Mermin and Svetlichny parameters in terms of measurable quantities is nontrivial here. However, guided by the previous work \cite{alok2016quantum}, the measures of quantum correlations viz. Bell-CHSH inequality, teleportation fidelity and geometric discord have been expressed in terms of survival and transition probabilities for two flavour neutrino-system. It could be envisaged that such an exercise, though complicated, could be carried out for the three flavour case.

\section{Conclusion}\label{conclusion}

Different facets of nonclassicality have been investigated for the neutrino system by considering the three flavour scenario of neutrino oscillation. The matter effects are included in order to carry out the analysis in the  context of the ongoing neutrino experimen NO$\nu$A and T2K and also for the future experiment DUNE. The analysis is carried out by considering both neutrino and antineutrino beams for the experiments.  The quantum correlations show sensitivity to the neutrino mass hierarchy, i.e. the sign of $\Delta_{31}$. It is a general feature displayed by all the correlations that the sensitivity to the mass hierarchy becomes more prominent for the high energy and long baseline experiment like DUNE compared to NO$\nu$A and T2K experiments. The results also suggest that in order to probe the various measures of nonclassicality in neutrino sector, one must use neutrino beam for the positive sign of $\Delta_{31}$ and an antineutrino beam otherwise.
\section*{Acknowledgments}
We acknowledge useful discussions with B.C. Hiesmayr and G. Guarnieri. This work is partially supported by DST India-BMWfW Austria Project Based Personnel Exchange Programme for 2017-2018. SB acknowledges partial financial support from Project No. 03 (1369)/16/EMR-II, funded by the Councel of Scientific \& Industrial Research, New Delhi.
%
\end{document}